\documentclass{aa}  

\usepackage{graphicx}
\usepackage{amsmath}
\usepackage{txfonts}
\usepackage{float}
\usepackage{placeins}
\usepackage{stfloats}

\usepackage[varvw]{newtxmath}
\usepackage{xcolor}
\usepackage[normalem]{ulem}

\begin{document}

   \title{New self-consistent theoretical descriptions for mass-loss rates of O-type stars}

\author{        F. Figueroa-Tapia\inst{1}\thanks{Corresponding author: \texttt{felipe.figueroat@postgrado.uv.cl}}
        \and
            J. A. Panei\inst{4,}\inst{5}
    \and
            M. Curé\inst{1,}\inst{2}\thanks{email: \texttt{michel.cure@uv.cl}}
        \and
            I. Araya\inst{3}
    \and
            S. Ekström\inst{6}
    \and 
             A. C. Gormaz-Matamala\inst{7}
    \and 
             R. O. J. Venero\inst{4,}\inst{5}
    \and 
             L. S. Cidale\inst{4,}\inst{5}
}

\institute{Instituto de Física y Astronomía, Universidad de Valparaíso. Av. Gran Bretaña 1111, Casilla 5030, Valparaíso, Chile.
        \and
        Centro de Astrofísica, Universidad de Valparaíso. Av. Gran Bretaña 1111, Casilla 5030, Valparaíso, Chile.
        \and
        Centro Multidisciplinario de Física, Vicerrectoría de Investigación, Universidad Mayor, 8580745 Santiago, Chile
        \and
        Departamento de Espectroscopía, Facultad de Ciencias Astronómicas y Geofísicas, Universidad Nacional de La Plata (UNLP), Paseo del Bosque S/N (B1900FWA), La Plata, Argentina.
        \and
        Instituto de Astrofísica de La Plata, CCT La Plata, CONICET-UNLP, Paseo del Bosque S/N (B1900FWA), La Plata, Argentina.
    \and
    Department of Astronomy, University of Geneva, Chemin Pegasi 51, 1290 Versoix, Switzerland
    \and
    Astronomical Institute, Czech Academy of Sciences, Fri\v{c}ova 298, CZ-251 65 Ond\v{r}ejov, Czech Republic
}

   \date{Received XXX; accepted XXX}

\abstract 
{Massive O-type stars lose a significant fraction of their mass through radiation-driven winds, a process that critically shapes their evolution and feedback into the interstellar medium. Accurate predictions of mass-loss rates ($\dot{M}$) are essential for models of stellar structure and population synthesis.}
{We computed wind parameters for O-type stars using a self-consistent approach that couples the hydrodynamics of the wind with detailed calculations of the line acceleration.\ This approach follows the theory of radiation-driven stellar winds and, thus, allows us to derive mass-loss rate distributions for different atomic configurations of the stellar flux.}
{We used the \textsc{tlusty} code for stellar atmosphere models to compute tailored non-local thermodynamic equilibrium models; these models served as input radiation fields for the calculation of the force multiplier factor and the line-force parameters, for which we used the \textsc{locus} code. These line-force parameters were then iteratively coupled with the \textsc{hydwind} code to solve the wind hydrodynamics. The procedure was repeated until convergence and applied across a grid of stellar parameters for three chemical configurations.}
{We obtain self-consistent wind parameters for a broad set of O-type stellar models. The results show a systematic decrease in mass-loss rates with the inclusion of more elements in the radiation field, which is attributed to a strong effect on the UV region of the spectral energy distribution. As more elements are included, resulting in a larger number of spectral lines, the contribution from the UV diminishes, leading to lower mass-loss rates. We fitted three theoretical prescriptions for $\dot{M}$ using a Bayesian approach; this yielded Pearson correlation values greater than $0.92$ for all three model grids. It also allowed for the estimation of the wind momentum-luminosity relationships for each of the grids, yielding results similar to those based on observations of O-type stars.}
{}

               
\titlerunning{New self-consistent theoretical descriptions for mass-loss rates of O-type stars}
\authorrunning{F. Figueroa-Tapia et al.} 

   \maketitle

\section{Introduction}

Massive stars are pivotal constituents in the evolutionary history of galaxies due to their profound influence on both the dynamical and chemical aspects of the interstellar medium. Their life cycles are characterised by energetic feedback mechanisms, intense radiation, strong stellar winds, and explosive phenomena such as core-collapse supernovae, which contribute to the dispersal of heavy elements and the formation of compact objects, including neutron stars and black holes \citep{dedonder2003, goswami2021}. Despite their rarity when compared to lower-mass stars, their cumulative impact extends far beyond mere statistics, shaping star formation and influencing the evolution of their host galaxies \citep{vink2001}.

The evolutionary development of very massive stars is predominantly dictated by their ability to lose mass through powerful, radiation-driven winds \citep{vink2000, krticka2004}. These winds, especially in O-type stars, are propelled by the momentum transfer from photons to atomic ions, most efficiently through metal lines in the UV portion of the stellar spectrum \citep{lucy1970, lucy1971}. The resultant mass-loss rates not only substantially alter the surface chemical composition and observable wind parameters but also determine the evolutionary path, influencing whether the star will ultimately collapse into a neutron star or form a black hole \citep{romagnolo2024, hawcroft2024}. These processes are also instrumental in driving galactic chemical enrichment and regulating feedback cycles, which influence subsequent generations of star formation \citep{ibrahim2023}.

Accurate predictions of wind strengths and mass-loss rates are crucial not only for stellar evolutionary models but also for population synthesis efforts, supernova rate estimates, and understanding the propagation of energetic feedback at galactic scales. Yet, the theoretical modelling of line-driven stellar winds faces persistent challenges: traditional mass-loss prescriptions often depend heavily on empirical calibrations or assume oversimplified stellar atmospheres, potentially missing the subtleties of non-local thermodynamic equilibrium (NLTE) effects and the complex atomic physics governing line acceleration \citep{driessen2022}.

One of the dominant issues is the sensitive dependence of mass-loss rates on the atomic structure and chemical composition of the emergent stellar flux, particularly in the UV, where the bulk of momentum transfer occurs \citep{lucy1970, castor1974, vink2000, krticka2006}. Models that oversimplify this radiation field can yield mass-loss rates that systematically deviate from values estimated from observations, especially in the ‘weak wind’ regimes of O-type stars\citep{vink2001, hawcroft2024}. Recent observations and theoretical work suggest that metallicity, clumping, and the radial temperature profile of the wind play substantial roles in determining both the efficiency and the observable imprint of the wind. Addressing these complexities is crucial for reconciling theoretical models with observations and thereby enabling the reliable application of wind parameters to stellar populations in diverse environments \citep{sundqvist2018, vinks2021}.

The significance of improved wind models extends beyond stellar astrophysics. The relationship between stellar mass loss, wind momentum, and luminosity forms the empirical backbone for distance calibrations and extragalactic studies \citep{kudritzki1999, herrero2002}. Moreover, mass-loss rates are a key ingredient in simulations of galactic enrichment histories and re-ionisation \citep{meynet2002}, linking the fate of individual stars to the larger-scale evolution of the Universe.

In light of these challenges, there is a compelling motivation to develop wind models that are both physically consistent and broadly applicable. Recent advances in stellar atmosphere modelling, particularly NLTE codes and self-consistent iterative approaches, enable refined descriptions of the line-force parameters that drive wind acceleration \citep{gormaz2019, gormaz2022}. By systematically coupling hydrodynamics to detailed radiative transfer solutions, these models can probe the sensitivity of the wind to atomic composition and flux structure, yielding insights into both mass-loss trends and the theoretical limitations of existing prescriptions.

The objective of the present study is to build upon these advances by implementing a self-consistent methodology that combines hydrodynamical modelling with the iterative calculation of line-force parameters for O-type stars with a variety of atomic configurations. The results presented herein can be used to make more reliable theoretical predictions for mass-loss rates, assess the limitations of previous models, and provide improved estimations for the wind momentum-luminosity relationships (WLRs), which are vital for both stellar and extragalactic research.

This paper is structured as follows: Section 2 provides an overview of the radiation-driven wind formalism and the details of the hydrodynamical approach. Section 3 outlines the methodology for model construction and the iterative determination of wind parameters. Section 4 presents the theoretical results, including the obtained distributions for the mass-loss rates and their corresponding Bayesian-fitted formulas, as well as a theoretical estimate of the WLR. Finally, Sect. 5 discusses the implications of the findings for massive star evolution and future research directions.

\section{Stellar wind hydrodynamical m-CAK formalism}

The theoretical description of radiation-driven winds in massive stars was first established by the pioneering work of \citet[hereafter CAK]{castor1975}, which, building on the formulation of \citet{lucy1970} and \citet{castor1974}, provided a framework to describe the total acceleration exerted by spectral lines on the wind. This theory was later refined by authors such as \citet{abbott1982}, who incorporated the effect of ionisation changes throughout the wind, and by \citet{pauldrach1986} and \citet{friend1986}, who introduced the finite-disk correction factor. These modifications gave rise to the modified CAK theory, or widely known as the m-CAK theory, which treats the stellar wind as a spherically symmetric outflow, accelerated by the transfer of momentum from stellar photons to ions via an ensemble of spectral lines. The hydrodynamical structure of the wind is governed by the mass, momentum, and energy conservation equations, which together determine the radial velocity and density profiles.

The one-dimensional stationary mass conservation equation is given by
\begin{equation}
\dot{M} = 4 \pi r^2 \rho(r) v(r)\, ,
\label{eq:continuity}
\end{equation}
\noindent where $\dot{M}$ is the mass-loss rate, $r$ the radial coordinate in spherical geometry, $\rho(r)$ the mass density, and $v(r)$ the radial velocity of the wind. The equation of momentum can be expressed as
\begin{equation}
v \frac{dv}{dr} = -\frac{1}{\rho} \frac{dP}{dr} - \frac{GM_*(1 - \Gamma_\text{e})}{r^2} + g_\text{line}\left(r, v, \frac{dv}{dr}\right),
\label{eq:momentum}
\end{equation}

\noindent where $P$ is the gas pressure, $M_*$ the stellar mass, $\Gamma_\text{e}$ the Eddington factor due to electron scattering, and $g_\text{line}$ is the radiative line acceleration. The term $(1 - \Gamma_\text{e})$ reflects the reduction of the effective gravitational force by radiative acceleration from Thomson scattering.

The total line acceleration ($g_\text{line}$) is expressed in terms of the electron scattering acceleration ($g_\text{es}$) and the so-called force multiplier, $\mathcal{M}(t)$:
\begin{equation}
g_\text{line} = \frac{\sigma_\text{e} F}{c} \mathcal{M}(t) = g_\text{es} \mathcal{M}(t),
\label{eq:gline}
\end{equation}

\noindent where $\sigma_\text{e}$ is the Thomson scattering cross-section, $F$ the stellar flux integrated over frequency, and $c$ the speed of light. The force multiplier quantifies the enhancement in radiative acceleration resulting from metal lines, relative to pure electron scattering.

The general expression for $\mathcal{M}(t)$ is\begin{equation}
\mathcal{M}(t) = \frac{g_\text{line}}{g_\text{es}} = \sum_\text{lines} \frac{\Delta \nu_D F_\nu}{F}  \frac{1 - e^{-\eta t}}{t}\, ,
\label{eq:forcemultiplier}
\end{equation}

\noindent where $\Delta \nu_D = v_\text{th}/\lambda$ is the Doppler width due to thermal motions, $v_\text{th}$ is the thermal velocity, and $F_\nu$ the emergent monochromatic flux at frequency $\nu$. The quantities $t$ and $\eta$ encode the optical depth parameter for an expanding atmosphere and the ratio of line to electron scattering opacity, respectively.

The local optical depth parameter for a comoving frame, $t$, is defined as
\begin{equation}
t = \sigma_\text{e} \ v_\text{th} \ \rho(r) \ \left( \frac{dv}{dr} \right)^{-1},
\label{eq:opticaldepth}
\end{equation}
\noindent while $\eta$ is given by
\begin{equation}
\eta = \frac{\pi e^2}{m_e c} \, g_l f_l \left( \frac{N_l/g_l - N_u/g_u}{\rho(r) \sigma_\text{e} \Delta \nu_D} \right).
\label{eq:eta}
\end{equation}
\noindent Here $e$ is the electron charge, $m_e$ the electron mass, $f_l$ the oscillator strength, and $N_l, N_u$ and $g_l, g_u$ are the atomic populations and statistical weights of the lower and upper levels, respectively.

For practical applications, the total effect of line-acceleration is often expressed through a parameterised form of the force multiplier as a power law:
\begin{equation}
\mathcal{M}(t) = k \ t^{-\alpha} \left( \frac{N_{e, 11}}{W(r)} \right)^\delta,
\label{eq:forcemultipliercak}
\end{equation}

\noindent where $k$, $\alpha$, and $\delta$ are the line-force parameters. Here, $N_{e, 11}$ is the electron number density in units of $10^{11}$\,cm$^{-3}$ and $W(r)$ is the dilution factor. The parameter $k$ represents the effective number of lines contributing to the acceleration, $\alpha$ represents the proportion of the line-force generated by optically thick lines relative to the total line-force, which governs the influence of both optically thick and thin lines, and $\delta$ accounts for the change of ionisation throughout the wind. 

The dilution factor ($W(r)$) accounts for the geometric attenuation of the radiation field from a star of radius $R_*$, and is defined as
\begin{equation}
W(r) = \frac{1}{2} \left( 1 - \sqrt{1 - \left( \frac{R_*}{r} \right)^2 } \right),
\label{eq:dilutionfactor}
\end{equation}

\begin{figure}
\centering
\includegraphics[width=\linewidth]{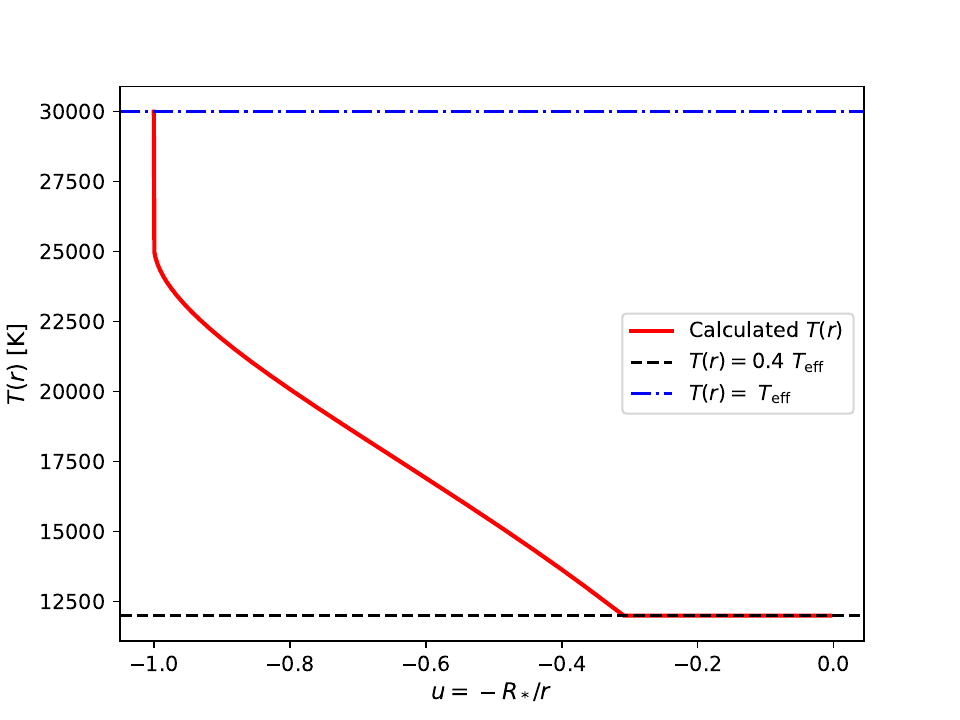}
\caption{Comparison of the wind temperature profiles, $T(r)$, commonly used in the atomic calculations of the force multiplier factor (Eq.~\ref{eq:forcemultiplier}). As can be seen, it is important to approximate the temperature profile throughout the wind, given its variation along the radial coordinate, rather than relying on the two fixed values typically adopted in the isothermal treatment.}
\label{fig:temp_prof}
\end{figure}

To improve the accuracy of the computed radiative acceleration and mass-loss rates, this work builds on the methodology introduced by \citet{gormaz2019}, which incorporated a more realistic description of the ionisation balance and level populations in stellar winds. In that approach, the evaluation of the atomic partition function, $U_i(T),$ included the temperature dependence derived by \citet{cardona2010}:

\begin{equation}
    U_i(T) = U_{i, 0} + G_{jk} e^{-\epsilon_{jk}/T} + \frac{m}{3} (n_*^3-343)e^{-\hat{E}_{n_*jk}/T},
    \label{eq:cardona}
\end{equation}

\noindent where $\hat{E}_{njk}$ is the mean excitation energy of the last level of the ion, and $n_*$ is the maximum excitation stage considered. $G_{jk}$, $\epsilon_{jk}$, and $m$ are tabulated values from \cite{cardona2010}; for non-tabulated ions, the prescription of \cite{Lattimer2021ApJ...910...48L} is adopted.

Because the winds of massive stars deviate from local thermodynamic equilibrium (LTE), \citet{gormaz2019} applied the quasi-NLTE corrections proposed by \citet{puls2005} to modify the ionisation balance as

\begin{equation}
\left( \frac{N_{i+1}}{N_i} \right)_{\mathrm{NLTE}} \approx \bigg(\frac{N{\mathrm{e}}}{W}\bigg)^{-1} \left( \zeta_i + W \left( 1 - \zeta_i \right) \right) \sqrt{\frac{T_{\mathrm{e}}}{T_{\mathrm{R}}}} \left( \frac{N_{i+1} N_{\mathrm{e}}}{N_i} \right)_{\mathrm{LTE}},
\label{eq:quasinlte}
\end{equation}

\noindent where $\zeta_i$ represents the fraction of recombination processes that go directly to the ground stage, $T_e$, the electron temperature, and $T_R$ the radiative temperature. Furthermore, to compute the fractional population of ions in stage $i$ ($X_i$), the formalism of \citet{abbott1985} was implemented, distinguishing between metastable and non-metastable states:

\begin{equation}
        \left(\frac{X_i}{X_1}\right) = \left\{
        \begin{array}{ll}
                \left(\frac{X_i}{X_1}\right)_{\mathrm{LTE}} & \text{metastable levels} \\\\
                W(r)\left(\frac{X_i}{X_1}\right)_{\mathrm{LTE}} & \text{others.}
        \end{array}
        \right.
        \label{eq:metalevels}
\end{equation}

\noindent This treatment of ionisation and excitation, combining the partition function, quasi-NLTE corrections, and the explicit consideration of metastable levels, leads to a more realistic estimation of the line-force parameters $(k, \alpha, \delta)$, following the procedure proposed in \cite{gormaz2019}.

The subsequent refinement introduced by \citet{gormaz2022} further enhanced the model by incorporating a non-isothermal radial temperature profile $T(r)$ for computing the local thermal velocity $v_\text{th}$ and level populations under realistic atmospheric conditions. The adopted description, following \citet{sundqvist2019}, is given by

\begin{equation}
T(r) = T_\text{eff} \bigg( W(r) + \frac{3}{4} \tau_F \bigg)^{1/4}\,,
\label{eq:temperature_profile}
\end{equation}

\noindent where $\tau_F$ is the flux-weighted optical depth,

\begin{equation}
    \tau_F (r) = \int \rho(r') \ \kappa_F(r') \bigg( \frac{R_*}{r'} \bigg)^2 \ dr'\,,
\end{equation}

\noindent and $\kappa_F$ is the opacity in the co-moving frame.

This temperature profile, consistent with radiative equilibrium in extended atmospheres, captures the temperature decline with radius more accurately than isothermal assumptions. To prevent unphysical cooling in the outermost layers, a minimum temperature of $0.4 \ T_\text{eff}$ was imposed, following \citet{gormaz2022}, as illustrated in Fig. \ref{fig:temp_prof}.

Altogether, the use of a non-isothermal temperature structure, a temperature-dependent partition function, and a quasi-NLTE treatment of ionisation provides a more realistic description of the physical conditions in hot star winds. These approximations allow the local ionisation balance and radiative acceleration to respond consistently to changes in temperature and density, capturing the coupling between the radiation field and the outflowing gas more accurately. As a result, the computed line-force parameters should reflect more realistic conditions in the wind rather than idealised LTE or isothermal assumptions, leading to more reliable estimates of the mass-loss rates and terminal velocities. This level of physical consistency is particularly important when comparing models across different stellar parameters and chemical compositions, as it ensures that the derived trends are the outcome of genuine radiative–hydrodynamical effects rather than artefacts of simplified assumptions.

\section{Self-consistent methodology and numerical calculations}

To derive the wind structure from a set of stellar parameters, a self-consistent methodology is adopted, in which the hydrodynamical profiles of the wind and the line-force parameters are determined iteratively from the stellar parameters and the radiation field. This approach is based on the procedure introduced by \citet{gormaz2019,gormaz2022} and further extended in this work to cover a larger parameter space. The method involves three core components: calculating the stellar radiative flux, solving the hydrodynamical wind equations iteratively coupled to the line acceleration, and constructing self-consistent grids of wind models.

\subsection{Radiative field calculation}

\begin{figure*}
\sidecaption
\centering
    \includegraphics[width=12cm]{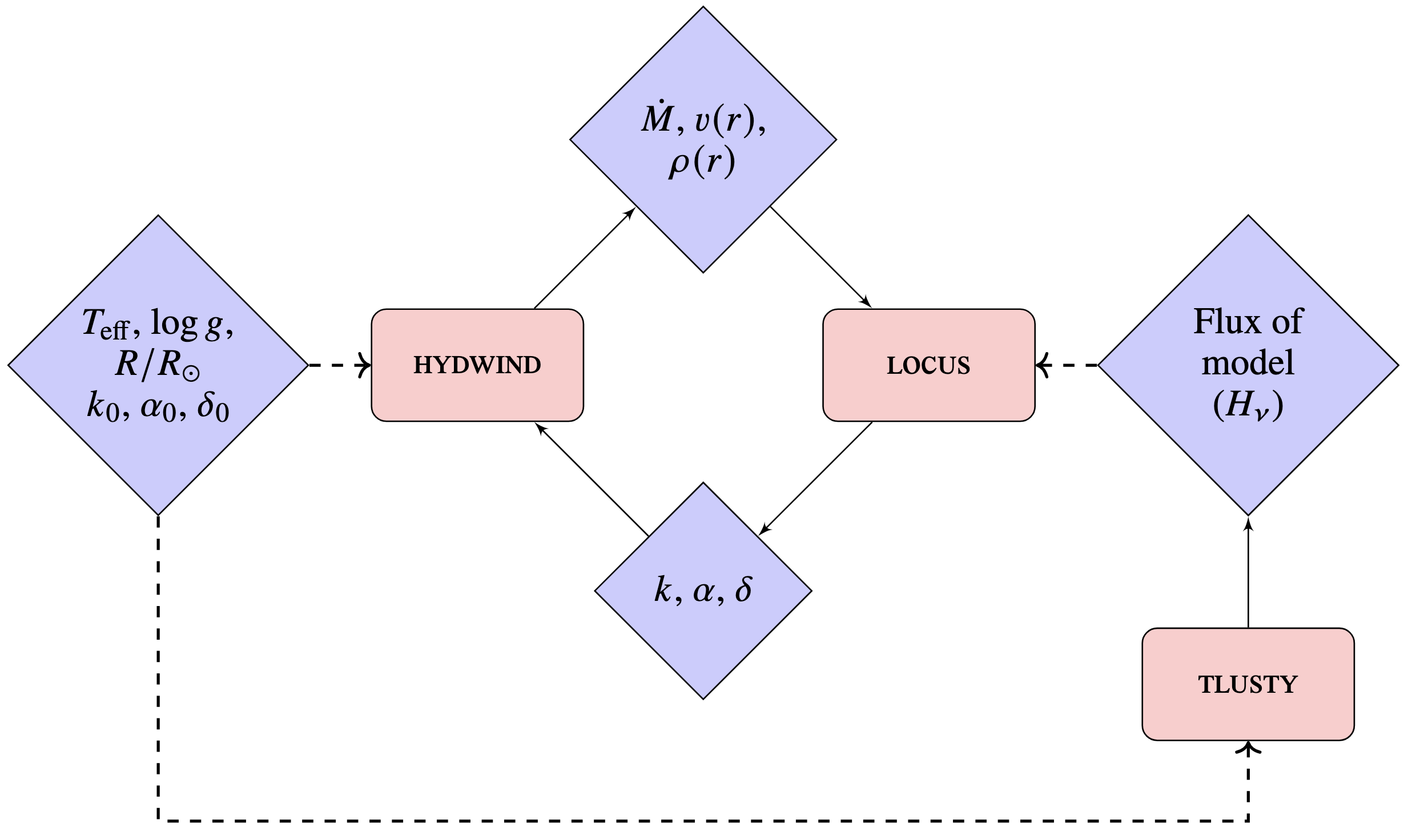}
    \caption{Iterative process introduced by \cite{gormaz2019} combining the codes \textsc{locus}, \textsc{hydwind}, and \textsc{Tlusty}. This process describes the calculation of the line-force and wind parameters from a unique set of stellar parameters. After computing the new values of $k$, $\alpha$, and $\delta$, they are compared with the previous ones; if Eq. \ref{eq:convergence} is satisfied, the process terminates.}
    \label{fig:locusloop} 
\end{figure*}

The efficiency of radiation-driven winds in hot stars is strongly determined by the structure of the emergent stellar flux, particularly in the UV range, where most of the momentum transfer occurs. To capture this dependence, a set of stellar atmosphere models was computed using the \textsc{Tlusty} code \citep{hubeny1995}, which solves the radiative transfer and statistical equilibrium equations under NLTE and hydrostatic assumptions in plane-parallel geometry, covering a wavelength range from $100$ \text{\r{A}} to $300$ $\mu$m. 

Each flux model was constructed for a specific combination of $T_\text{eff}$ and $\log g$. Assuming solar metallicity, the models were constructed with varying levels of atomic complexity, including more lines, allowing for a controlled analysis of how the inclusion of different chemical species affects the radiative driving, while ensuring that models with the same parameters from different grids have identical total fluxes.

For each atomic configuration, \textsc{Tlusty} was executed independently, ensuring proper convergence of the level populations under NLTE conditions before extracting the emergent flux. These spectra reflect the cumulative effect of increasing atomic opacity on the stellar radiation field.

The resulting fluxes, which account only for the photospheric layer of the star, were used as input for calculating the line acceleration in the wind. By systematically controlling the atomic content of the radiation field, this approach enables a detailed investigation of how the number of spectral lines contributes to the overall driving mechanism and influences the derived mass-loss rates and wind parameters.

\subsection{Iterative self-consistent procedure}

The m-CAK procedure requires a coupled treatment of hydrodynamics and line-force calculations. A self-consistent solution is achieved by iterating between the velocity and density profiles of the wind and the determination of the line-force parameters ($k$, $\alpha$, $\delta$).

We adopted the following iterative scheme, seen in Fig. \ref{fig:locusloop}. Starting from a randomly selected set of line-force parameters, consistent with previous ranges found in the literature (e.g. \citealt{kudritzki2002}; \citealt{noebauer2015}; \citealt{gormaz2019, gormaz2022}), an initial hydrodynamic stationary solution is computed with the \textsc{hydwind} code \citep{cure2004}. This provides velocity and density profiles together with preliminary estimates of the mass-loss rate and terminal velocity. Using these profiles, the newly developed \textsc{locus} code, based on the work of \citet{gormaz2019}, calculates the force multiplier M(t) based on three electron density profiles. Crucially, unlike the incident radiation field, the wind material retains a full solar chemical composition, including metals up to iron; this composition is adopted from the complete CMFGEN atomic database \citep{hillier1990a, hillier1998} and is based on the m-CAK formalism.

From the resulting $\mathcal{M}(t)$, the line-force parameters are obtained via a linear fit via Eq.~\ref{eq:forcemultipliercak}. The updated set of parameters is then fed back into \textsc{hydwind} to compute an improved hydrodynamical solution. This iteration scheme is repeated as seen in Fig. \ref{fig:locusloop}, until convergence is reached. Convergence is defined when the relative change of each line-force parameter between two iterations satisfies

\begin{equation}
\left| \frac{p_i - p_{i-1}}{p_i} \right| \leq 5 \times 10^{-4},
\label{eq:convergence}
\end{equation}

\noindent where $p_i$ denotes any of the parameters ($k$, $\alpha$, $\delta$) computed in the $i$-th iteration, and $p_\text{calculated}$ represents the value of the parameters obtained before applying the damping factor. In this way, we guarantee four significant digits of precision. In practice, convergence is typically achieved within four to five iterations. To avoid oscillatory behaviour between two solutions that are not reducing their distance, we introduced a damping factor ($j_{\mathrm{eff}} = 0.98$) to smooth parameter jumps between successive iterations, such that each iteration is defined as

\begin{equation}
    p_{i+1} = p_i - (p_i - p_\text{calculated}) \times j_\text{eff}^{i-1} \, .
\end{equation}

Finally, a maximum of 70 iterations was imposed due to oscillations between nearby solutions that failed to converge, although this limit was rarely reached. This procedure ensures that the final values of $k$, $\alpha$, and $\delta$ are uniquely determined by the stellar parameters and consistent with the hydrodynamical structure of the wind, reducing the number of free parameters in the m-CAK framework.

\subsection{Building the self-consistent grids}

To investigate the global behaviour of O-type stellar winds, we constructed large grids of self-consistent models spanning the typical range of stellar parameters for Galactic O stars. The grids cover non-rotating models with effective temperatures between $30000$ and $50000$ K in steps of $1000$ K, surface gravities in the range $\log g = 2.9-4.3$ each $0.2$ [dex], and stellar radii from $7$ to $70$ $R_*/R_\odot$, with a $1.5$ step. These ranges were chosen because the gas in the stellar envelope can be reasonably assumed to be fully ionised, allowing the electron scattering opacity to remain approximately constant throughout the atmosphere \citep{bestenlehner20}. This condition ensures the validity of the radiative assumptions on which the m-CAK formalism is based.

These intervals sample both the main sequence, but also the giant and supergiant stages, ensuring that the results can be directly applied to observed populations in the HR diagram. For each set of stellar parameters, the stellar mass and luminosity were derived from evolutionary relations, providing the fundamental input for the hydrodynamical calculations.

The iterative self-consistent procedure described in the previous subsection was applied to each stellar model until convergence was achieved. The resulting line-force parameters ($k$, $\alpha$, $\delta$), mass-loss rate, and terminal velocity were then stored in the grids.

Since our goal is to quantify how sensitive the force multiplier is to the adopted radiation field, we constructed three independent grids of self-consistent wind models, each based on fluxes calculated with \textsc{Tlusty}. These grids correspond to the following prescriptions:  
\begin{enumerate}
    \item {H grid}: radiation field including only hydrogen, serving as the simplest reference case. 
    \item {HHe grid}: radiation field including hydrogen and helium. 
    \item {CNO grid}: radiation field including hydrogen, helium, carbon, nitrogen, and oxygen, thereby incorporating the first relevant metal lines responsible for driving the wind in the UV.  
\end{enumerate}

This stepwise inclusion of atomic species allows us to directly assess the impact of line opacity on the force multiplier and wind parameters. By comparing the three grids, we can quantify how progressively adding helium and metal ions modifies the derived values of $k$, $\alpha$, and $\delta$, and consequently the mass-loss rates and terminal velocities.

However, not all parameter combinations converged to physically realistic winds, and there may be numerical solutions that represent conditions that cannot be realistic. Models were rejected if (i) they failed to converge numerically after 70 iterations, (ii) the velocity profile obtained unrealistic values ($v_\infty \geq 3000$ km s$^{-1}$), (iii) the resulting mass from the combination of the selected $\log g$ and $R_*$ was higher than $120$ $M_\odot$, or (iv) the force multiplier model showed a linear fit quality below $R^2 = 0.98$. After this filtering process, the final grids retained several hundred models per flux prescription. Table \ref{tab:grids} summarises the extent of the final grids.

\begin{table}
\caption{Details of the selected models for the three self-consistent grids. \label{tab:method:2:grids}}            
\label{tab:grids}      
\centering                                     
\begin{tabular}{r r r}       
\hline\hline                        
{Grid} & {Converged Models}  & {Selected Models} \\
\hline                                  
        H       &  4716                         & 1438  \\
        HHe     &  5155                         & 1069  \\
        CNO &  4968                     & 707   \\
\hline                                          
\end{tabular}
\tablefoot{Despite the significant loss of models due to the convergence and selection criteria, the resulting sample remains sufficiently large for subsequent analysis of wind parameters.}
\end{table}

\section{Results}

\subsection{Mass-loss rate distributions}

\begin{figure}
        \centering
        \includegraphics[width=\linewidth]{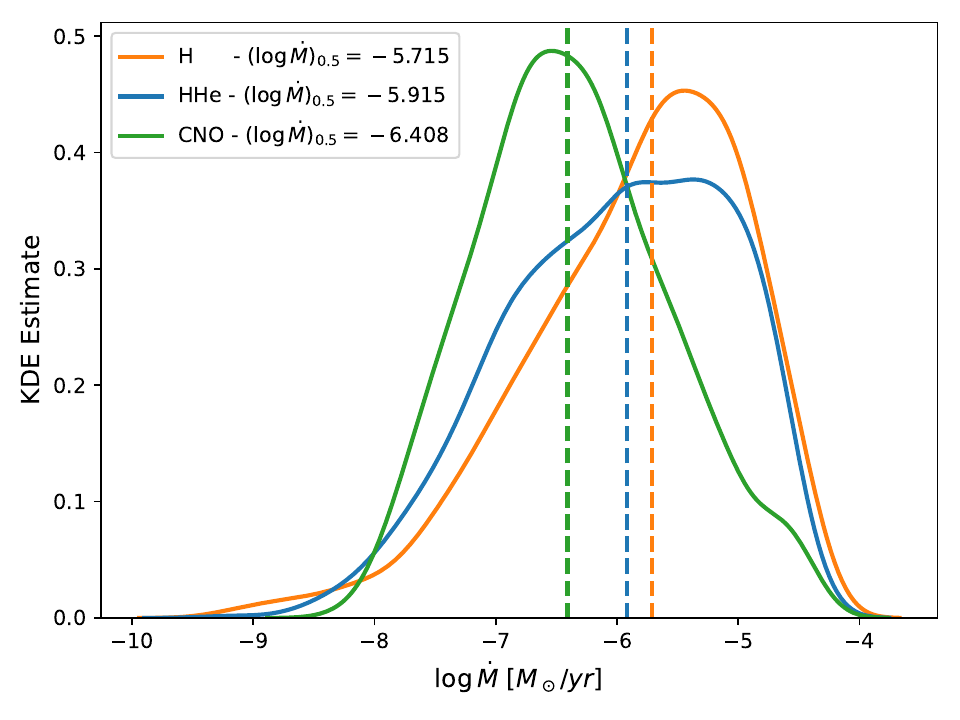}
        \caption{$\dot{M}$ kernel density estimation distribution for each grid. The median values of each distribution are indicated by vertical dashed lines. A clear trend emerges, showing lower mass-loss rates for grids with more complex radiation fields. This indicates that simpler radiative configurations yield systematically higher mass-loss rates, while the inclusion of more atomic transitions results in predictions consistent with the low mass-loss regime.\label{fig:mdotdist}}
\end{figure}

The data from the filtered grids in Fig. \ref{fig:mdotdist} reveal that the mass-loss rate is significantly influenced by the number of elements included when calculating the force multiplier, $\mathcal{M}(t)$. This relationship is described through Eq. \ref{eq:forcemultiplier}, which demonstrates a direct dependence on $F_\nu$. As more elements are incorporated, the frequency-dependent flux ($F_\nu$) is distributed over a larger number of transitions, effectively diluting the impact of the flux value. However, the total flux ($F$) does not diminish enough to equal the spectral flux density contributions of these additional lines in the overall calculation. This systematic reduction in mass-loss rates highlights the sensitivity of $\dot{M}$ to the complexity of the radiation field. It also emphasises the importance of accurately accounting for the number of atomic lines, as neglecting this effect can lead to systematically higher mass-loss rates, particularly in atomically simpler grids dominated by hydrogen lines. This trend highlights the significance of detailed radiative transfer calculations in modelling stellar winds in massive stars.

Incorporating the findings from the three grids reveals a clear trend in mass-loss rate distributions, as shown in Fig. \ref{fig:mdotdist}. This trend has significant implications as it naturally yields lower mass-loss rates, aligning more closely with observational constraints that suggest values below traditional theoretical predictions \citep{bouret2012, surlan2013, vink2022}, by increasing the complexity of the radiation field used in calculations. The H grid, for example, shows stronger winds due to its higher median mass-loss rates compared to the other grids. This grid also has the fewest wind models with low $\dot{M}$, with mass-loss rates falling below those of the HHe and CNO grids. 

\begin{figure}
\centering
\includegraphics[width=\linewidth]{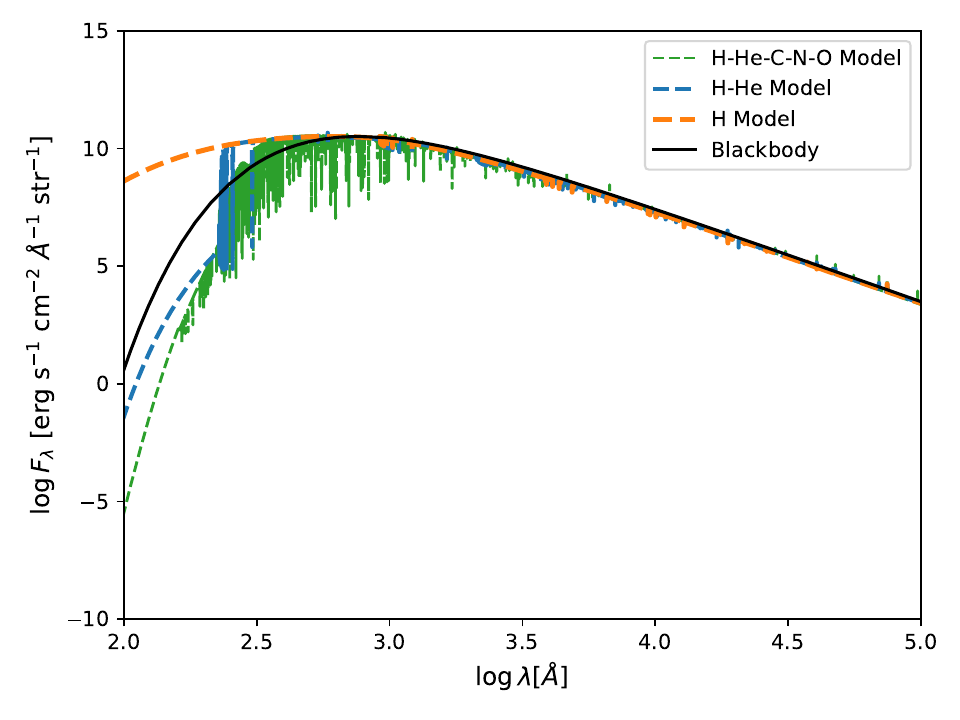}
\caption{Emergent fluxes computed with \textsc{Tlusty} for a representative stellar model ($T_\mathrm{eff} = 38000 \ \mathrm{K}$ - $\log g = 3.4$). Each curve corresponds to a different radiative configuration, illustrating the effect of increasing atomic complexity. As can be seen, the effect, especially in the UV region, is considerable and will translate into an effect on the value of the mass-loss rate.}
\label{fig:flux_composition}
\end{figure}

In contrast, the HHe grid demonstrates a slight reduction in mass-loss values, with fewer models exhibiting rates above $10^{-6} M_\odot / \text{yr}$ and an excess of models near $10^{-7} M_\odot / \text{yr}$. While the median mass-loss rate for the HHe grid is similar to the H grid, this is expected given the modest difference in their radiation fields (see Fig. \ref{fig:flux_composition}). Including helium introduces additional lines but does not significantly alter the flux compared to the CNO case. The UV region, crucial for line-acceleration \citep{lucy1970}, is also more affected in the HHe grid than in the H grid, resulting in a slight shift in the mass-loss distribution.

The most significant changes occur in the CNO grid, which shows the most significant reduction in mass-loss rates. Following the trend observed in the HHe grid relative to the H grid, the CNO grid features an even greater excess of models with mass-loss rates closer to $10^{-7} M_\odot / \text{yr}$. This grid has the most extensive inclusion of lines and shows the most substantial reduction in flux in the UV region (see Fig. \ref{fig:flux_composition}). Consequently, the number of models with high mass-loss rates dramatically decreases compared to the H grid. This result aligns well with expectations, as it yields the lowest mass-loss rates, effectively addressing the previously mentioned issue of overestimation. Moreover, this systematic reduction in mass-loss rates underlines the importance of accounting for heavier elements in radiation-driven wind models. Heavier elements contribute significantly to line opacity, which can alter the dynamics of line-acceleration. The results suggest that models considering a broader range of elements can better match observational constraints and theoretical predictions for lower mass-loss rates. This is particularly relevant when studying stars with weaker winds, where discrepancies between theoretical and observed rates are more pronounced \citep{martins2005}.

\subsection{Theoretical linear descriptions}

The large number of theoretical models employed in this work provides the statistical leverage necessary to construct reliable prescriptions for the mass-loss rate ($\dot{M}$) directly from stellar parameters. The extended dataset enables us to quantify systematic trends with improved precision, reducing the impact of sample variance compared to previous studies that relied on smaller model grids, such as \cite{gormaz2019}. Within this framework, our objective is to derive linear relations for $\dot{M}$ across the three self-consistent grids and to confront these relations with those reported in earlier works, thereby enabling both the validation of our approach and the identification of potential departures from established trends.

For this purpose, we adopted a Bayesian linear regression scheme implemented with the Bayesian software \textsc{stan} \citep{stan2024}. The Bayesian formalism ensures a consistent treatment of uncertainties, which we report as $2\sigma$ credibility intervals extracted from the posterior distribution of each parameter. This linear adjusted relationship is motivated by previous applications to stellar wind studies, such as \cite{vink2000}, \cite{kk18}, \cite{gormaz2019}, and \cite{bjorklund2022}, where the emphasis was placed on relying exclusively on intrinsic stellar parameters such as the effective temperature ($T_\text{eff}$ [K]), the surface gravity ($\log g$), and the photospheric radius ($R_*$ [$R_\odot$]). By excluding wind-derived quantities, the resulting predictions remain anchored strictly to stellar properties, which enhances their physical interpretability and facilitates comparison with observations.

This approach offers several key advantages. By relying only on stellar parameters, the resulting relationships for $\dot{M}$ are easier to interpret and more straightforward to use, especially when compared with observational data. This focus underscores the central role of intrinsic stellar properties in governing mass-loss rates, independent of the complexities introduced by wind dynamics. Consequently, this framework not only simplifies theoretical predictions but also facilitates alignment with observational studies, allowing for broader applicability. Nevertheless, it is essential to note a significant limitation of the current models: solar metallicity is assumed. As a result, the well-established dependence of mass-loss rates on metallicity \citep[Z;][]{puls2008, mokiem2008, vink2018} is not incorporated into our analyses. This omission constrains the generality of the findings, as variations in $Z$ are known to significantly influence wind dynamics and $\dot{M}$. Future expansions of the framework to include non-solar metallicities will be critical for achieving a more comprehensive understanding of mass-loss processes across diverse stellar environments.

\begin{figure}
        \centering
        \includegraphics[width=1\linewidth]{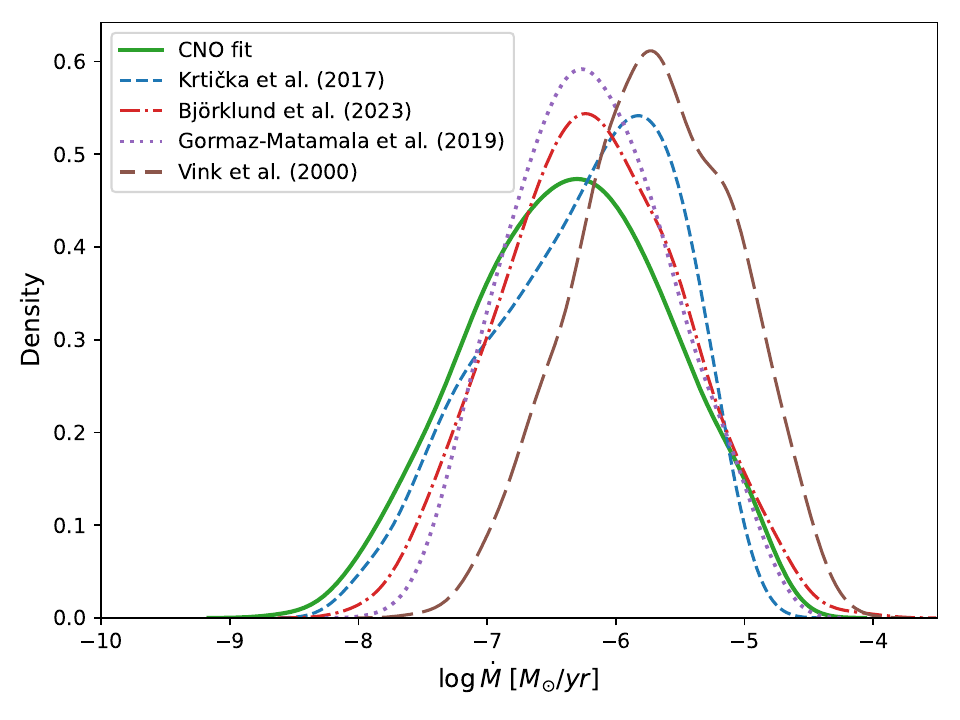}
        \caption{Comparison of mass-loss rates obtained from the adjusted theoretical description of the most complex grid in terms of atomic structure (CNO; solid green line) with results from other studies. As shown, the models derived in this work exhibit a better description of low mass-loss rates, particularly in regions where $\log \dot{M}$ values fall below predictions of previous studies, reaching values as low as $10^{-8}$ [$M_\odot$ / yr].
                \label{fig:mdotcomp}}   
\end{figure}

For the H grid, the model was successfully fitted, as illustrated in Fig. \ref{fig:hbayes}. This Bayesian fit demonstrates that most parameter distributions exhibit minimal correlation, except for $\log  T_\text{eff}$ and $\log g$, which display a slight correlation, and the free parameter of the relation $ \beta_0$. However, this relationship is not strong enough to significantly influence the calculations or compromise the model's reliability. The fit highlights a consistent performance across the parameter space, ensuring a robust interpretation of the results. With an $R^2 = 0.939$, the model provides a reliable representation of the underlying data trends. The derived relationship is expressed as follows:
\begin{align}
        \label{eq:h_fit}
        \begin{split}
                \log  \dot{M}_\text{H} \ =      {} & \ - \ 23.05 \ (\pm 0.37) \\
                {} & \ + \ \ \ 1.32 \ (\pm 0.06) \ \log  \ (R_*/R_\odot) \\
                {} & \ - \ \ \ 3.36 \ (\pm 0.07) \ \log  g \\
                {} & \ + \ 17.55 \ (\pm 0.29) \ \log  \ (T_{\rm eff}/1000 \ \rm K) 
        \end{split},
\end{align}

\noindent where $g$ is in units of [cm/s$^{2}$].

In the case of the HHe grid, the parameters undergo significant changes due to the noticeable variations in the flux (see Fig. \ref{fig:flux_composition}). Similar to the H grid, the Bayesian fit results show promising outcomes, as illustrated in Fig. \ref{fig:hhebayes}, where correlations remain low, except for the same cases discussed previously. This consistency reinforces the validity of the modelling approach across different grids. The dispersion of the fitted models decreases slightly compared to the H grid, indicating a minor improvement in the model's predictive precision. Furthermore, the fit achieves an $R^2 = 0.946$, reflecting a high-quality representation of the data and the model's ability to capture the underlying trends effectively. The adjusted formula can be written as
\begin{align}
        \label{eq:hhe_fit}
        \begin{split}
                \log   \dot{M}_\text{HHe} \ = {} & \ - \ 32.51 \ (\pm 0.44) \\
                {} & \ + \ \ \ 1.34 \ (\pm 0.07) \ \log  \ (R_*/R_\odot) \\
                {} & \ - \ \ \ 4.21 \ (\pm 0.12) \ \log  g \\
                {} & \ + \ 25.22 \ (\pm 0.43) \ \log  \ (T_{\rm eff}/1000 \ \rm K) 
        \end{split}.
\end{align}

Finally, for the CNO grid, a noticeable change is observed in the adjusted parameters. As the most complex grid and with the largest variation in the UV region of the radiative fluxes used, this outcome is expected. Upon reviewing Fig. \ref{fig:cnobayes}, it is evident that the same trends previously described persist, with low correlations, except for the same parameters mentioned earlier. This stability across grids indicates that the overall methodology is robust despite the increasing complexity of the models. Additionally, the dispersion shows a slight decrease compared to the HHe grid, although this reduction is almost negligible. Consequently, the adjusted model presents an $R^2 = 0.929$, and its formula can be written as
\begin{align}
        \label{eq:cno_fit}
        \begin{split}
                \log   \dot{M}_\text{CNO} \ =      {} & \ - \ 20.74 \ (\pm 0.61) \\
                {} & \ + \ \ \ 1.44 \ (\pm 0.09) \ \log  \ (R_*/R_\odot) \\
                {} & \ - \ \ \ 3.85 \ (\pm 0.15) \ \log  g \\
                {} & \ + \ 16.81 \ (\pm 0.45) \ \log  \ (T_{\rm eff}/1000 \ \rm K) 
        \end{split}.
\end{align}

As shown in Fig. \ref{fig:mdotcomp}, when comparing the adjusted prescription for the most complex model (and thus the one predicting the lowest mass-loss rates) with other studies that describe winds based on stellar parameters, our results extend the predicted range of mass-loss rates towards systematically lower values within the explored parameter space. The relation proposed by \citet{vink2000} predicts higher mass-loss rates, showing a peak in the distribution above $10^{-6}$ [$M_\odot$ / yr] and failing to reproduce winds below $10^{-7}$ [$M_\odot$ / yr]. In contrast, the models by \citet{gormaz2019} and \citet{bjorklund2022} provide results that are more consistent with the higher end of the distribution, although they still deviate at lower $\dot{M}$ values, albeit not drastically. The adjusted relation derived in Eq.~\ref{eq:cno_fit}, however, succeeds in reproducing the lower mass-loss distribution wing with a similar trend to that of \citet{kk18}. Unlike the other distributions, the adjusted relation gives a more extensive representation of lower mass-loss rates, extending the range to values below $\dot{M}\sim10^{-8}$ [$M_\odot$ / yr] while remaining consistent with results for higher mass-loss rates.

\subsection{Estimation and comparison of the wind momentum-luminosity relationship}

One of the key outcomes of the m-CAK theory is its ability to describe stellar winds and obtain hydrodynamic solutions. This theoretical framework carries implications that merit observational calibration. Among the theoretical outcomes of radiation-driven wind theory is the WLR, which is expected to hold most clearly in the O supergiant regime under the assumption of a nearly constant CAK force-multiplier parameter $\alpha \sim 2/3$ \citep{kudritzki1999}. This relationship establishes a direct connection between the wind momentum, a measure of its strength and extent, and the stellar luminosity. Its significance lies in providing a physical link between the observable properties of stellar winds and the intrinsic characteristics of stars \citep{puls2008}. Quantitatively, this correlation implies that the wind momentum scales as a power law of the emitted radiation, expressed as

\begin{table*}[h!]
\caption{Fitted values for the WLR based on the data computed for each grid.\label{tab:wmlr_table}}  
\centering                                    
        \begin{tabular}{l l l l l}
                \hline\hline
                \multicolumn{1}{c}{Type of Models} & \multicolumn{1}{c}{$D_0$} & \multicolumn{1}{c}{$x$} & \multicolumn{1}{c}{$\alpha_\text{eff}$} & \multicolumn{1}{c}{$R^2$}\\
                \hline
                H                                               &       $18.40 \pm 0.01$               & $1.84 \pm 0.005$      & $0.543 \pm 0.003$ & $0.734$   \\
                HHe                                     &       $18.00 \pm 0.01$           & $1.85 \pm 0.005$      & $0.540 \pm 0.003$ & $0.728$   \\
                CNO                                     &       $18.50 \pm 0.01$           & $1.75 \pm 0.005$  & $0.571 \pm 0.003$ & $0.824$       \\
                \hline
                \cite{kudritzki1999b}   &       $20.40 \pm 0.85$            & $1.55 \pm 0.15$   & $0.650 \pm 0.060$      &       ----  \\
                \cite{vink2000}                 &       $18.68 \pm 0.26$            & $1.83 \pm 0.04$   & $0.548 \pm 0.012$      &       ----  \\
                \cite{herrero2002}              &       $19.27 \pm 1.82$            & $1.74 \pm 0.32$   & $0.580 \pm 0.120$      &       ----  \\
                \hline
        \end{tabular}

\tablefoot{A clear trend emerges, showing slight similarities across all grids, which vary primarily due to the dependence of $\dot{M}$ on the radiative field. Despite this variation, the obtained results align well with those reported by \cite{vink2000} and \cite{herrero2002}, demonstrating consistency with their observationally calibrated adjustments.}
\end{table*}

\begin{equation}
        \label{eq:wmlr_init}
        \dot{M} v_\infty  R_*^{1/2} \propto \ L^{1/\alpha_\text{eff}}.
\end{equation}

where $\dot{M}$ is the mass-loss rate, $v_\infty$ the terminal wind velocity, $R_*$ the stellar radius, and $L$ the luminosity. The effective exponent is defined as $\alpha_\text{eff} = \alpha - \delta \sim 2/3$, corresponding to the results obtained from the computed grids.

In other words, stellar luminosity can be derived from wind parameters that are measurable through observational means, offering a direct link between a star’s observable wind properties and its intrinsic characteristics. This makes the WLR a valuable tool for determining distances to remote celestial objects by studying their stellar winds.
Observational confirmations of this relationship have been achieved using stars from the Milky Way, the Magellanic Clouds, and other galaxies in the Local Group \citep{bresolin2004, evans2005, mokiem2005, garcia2025}, highlighting its universality.

Furthermore, studies such as those by \cite{puls1996} and \cite{kudritzki1999b} have shown that incorporating the $\beta$-law introduces variations in the WLR depending on the star's spectral type, luminosity class, and metallicity. These findings emphasise the complexity of the relationship and its sensitivity to different stellar properties. For instance, metal-rich environments tend to enhance the strength of stellar winds, resulting in adjustments to the WLR slope and intercept. This variability underscores the need for detailed calibrations of the relationship for diverse spectral types and metallicity environments. 

In this study, the comprehensive grid of theoretical models for O-type stars, spanning both main-sequence and supergiant phases, provides an opportunity to refine and better understand the WLR. By incorporating these models, we can identify the optimal parameters for the relationship, ensuring its applicability across a wide range of stellar conditions. Consequently, the equation initially described in Eq. \ref{eq:wmlr_init} can be reformulated as follows:

\begin{equation}
        \log  D = \log  D_0 + x \log  (L/L_\odot)\label{eq:wmlr_log},
\end{equation}

\noindent with $D = \dot{M} v_\infty (R/R_\odot)^{1/2}$ and $x = 1/\alpha_\text{eff}$. Expressing the WLR in this form simplifies the problem, as the parameters to calibrate are reduced to the slope and intercept of the linearised formula, making it more straightforward to fit model data effectively.

As shown in Table \ref{tab:wmlr_table}, the obtained results align well with general expectations, both in magnitude and overall behaviour. However, the grids presented here, particularly the more atomically complex ones, display a tendency to underestimate $\dot{M}$, resulting in values that fall below previous relationships. The H grid, characterised by its simpler atomic structure, exhibits a slightly steeper slope compared to the other grids, showcasing a resemblance to the results reported by \cite{vink2000}. This alignment suggests that the inclusion of only hydrogen in the radiative transfer calculations captures a trend consistent with theoretical predictions for stars with stronger winds.

The HHe grid, on the other hand, retains a slope nearly identical to that of the H grid. However, its intercept is reduced due to the lower mass-loss rates derived, which still maintain notable proximity to \cite{vink2000}. This subtle difference highlights the impact of helium on wind dynamics, where its inclusion alters the radiative driving without significantly changing the overall behaviour of the WLR.

Finally, the CNO grid demonstrates the shallowest slope among the three, aligning most closely with the results of \cite{herrero2002}. This grid not only presents the highest intercept of the three calculated grids, but also features the strongest correlation coefficient, $R^2 = 0.825$, reflecting a robust statistical agreement. Although the intercept for this grid is the highest among the three cases, it is crucial to consider that the variation in the slope (which depends on the line-force parameters) impacts the definition of the intercept. This occurs despite the tendency towards a lower value due to its lower mass-loss rates, as can be observed in Fig. \ref{fig:mdotdist}.

The results presented highlight how the WLR varies across grids with different atomic complexities. Each grid showcases distinct trends in slope and intercept, reflecting the influence of line-force parameters and wind properties on this relationship. Notably, the CNO grid achieves the highest correlation coefficient ($R^2 = 0.825$), demonstrating its strong alignment with theoretical expectations, while also exhibiting mass-loss rates and intercepts that are systematically lower in comparison with earlier prescriptions.

\section{Discussion and conclusions}

We employed the m-CAK prescription to self-consistently model stellar winds, aiming to improve upon the results obtained by \cite{gormaz2019}. To achieve this, three different grids of self-consistent winds were constructed, each containing hundreds of models. These grids explored variations in the atomic complexity of the radiative flux. This factor significantly impacted the mass-loss rate while having a relatively minor effect on the terminal velocity. This approach enabled a detailed examination of the influence of atomic parameters on wind properties, thereby contributing to a deeper understanding of the mechanisms driving stellar mass loss.

In this context, for the parameter $\dot{M}$, it was observed that as more atomic species were included in the force multiplier factor $\mathcal{M}(t)$ calculations (and consequently, a greater number of spectral lines were considered), the mass-loss rates decreased by up to a factor of $\sim 10$. The CNO grid yielded systematically lower mass-loss rate predictions, with the distribution peaking near $\sim 10^{-7}$ [$M_\odot$ yr$^{-1}$], whereas models that used the H grid overestimated the mass-loss rate compared with HHe and CNO grids, with a significantly higher peak at $\sim 10^{-6}$ [$M_\odot$ yr$^{-1}$] and a larger number of models around $\sim 10^{-5}$ [$M_\odot$ yr$^{-1}$]. This behaviour can be attributed to the effect of the radiation field on the calculation of $\mathcal{M}(t)$, which is highly sensitive to the UV region of the flux. As additional elements are introduced, the flux becomes increasingly affected, exhibiting a reduced UV contribution. This diminished UV flux weakens the radiative driving behind stellar winds, leading to a noticeable reduction in the mass-loss rate. Furthermore, adjustments were made using a formula similar to that of \cite{gormaz2019} but with a modified photospheric radius component. It yielded statistically robust results ($R^2 > 0.92$) consistent with current theoretical predictions for each grid and relies solely on stellar parameters as inputs. Consequently, the statistical robustness of the description increased due to the larger number of models included in the adjustment process.

The WLR was calibrated for the three grids, allowing a direct comparison with results from other authors who analysed O-type stars. The results from the calculation and fitting of this relationship are in good agreement with studies such as \cite{vink1999} and \cite{herrero2002}, displaying significant similarity in both slopes and intercepts for fast solutions. This indicates that a proper study of the WLR requires considering the largest possible number of elements contributing to the radiation field, as this strongly affects the computed mass-loss rates. Consequently, the WLR calibration experiences a shift towards lower intercept values, which subtly affects its fit. This effect was clearly reflected in the differences observed among the H, HHe, and CNO grids, illustrating how the inclusion of additional elements refines the relationship and leads to a better alignment between theoretical and observational results.

Overall, these results highlight the importance of adopting a self-consistent treatment of the radiation field when modelling stellar winds. The sensitivity of the mass-loss rate to the atomic composition of the flux emphasises that simplified prescriptions may exhibit a consistent trend towards higher values, whereas more comprehensive treatments yield values that show better agreement with observations. Thus, the methodology developed here represents a step forward in bridging theoretical predictions and empirical evidence, thereby strengthening the reliability of wind models for O-type stars.

These findings also open the door to comparisons with stellar evolution models, such as \textsc{Genec} \citep{meynet1994, ekstrom2008}, using observationally derived parameters. Indeed, the update of mass-loss rates has already had an impact on massive star evolution simulations \citep{gormaz2023,gormaz2024}, modifying, for example, the final stellar mass before the core-collapse \citep{romagnolo2024,gormaz2025}. The inclusion of the new mass-loss recipe presented here in its Bayesian form into \textsc{Genec} models and the study of the global impact it has on the evolution of massive stars will be the subject of a forthcoming paper. For future work, it is important to note that our models are currently restricted to solar metallicity and include elements only up to carbon. We plan to improve both aspects by employing flux grids with varying metallicities and by extending the calculations to include iron and other iron-group elements. Although their abundances are relatively low, these species contribute an enormous number of spectral lines in the far-UV, where radiative driving is most effective. Their inclusion is therefore expected to substantially modify the force multiplier parameters and yield more realistic mass-loss rates, providing a more complete and accurate description of the line-driving mechanism across different metallicities.

\begin{acknowledgements}
      FFT acknowledges support from ANID BECAS/DOCTORADO NACIONAL/2025-21250788. MC and IA thank the ANID FONDECYT project 1230131. MC also acknowledges support from the Centro de Astrofísica de Valparaíso (CAV), CIDI N. 21 (Universidad de Valparaíso, Chile). SE acknowledges support from the Swiss National Science Foundation (SNSF), grant number 212143. ACGM thanks the support from project 10108195 MERIT (MSCA-COFUND Horizon Europe). LC, JAP and ROJV acknowledge financial support from CONICET (PIP 11220200101337CO) and the Universidad Nacional de La Plata, Proyectos I+D 11/G192 (LC and ROJV) and 11/G193 (JAP and ROJV). Powered@NLHPC: This research was partially supported by the supercomputing infrastructure of the NLHPC (ECM-02). This project has received funding from the European Union (Project 101183150-OCEANS).
\end{acknowledgements}

\bibliographystyle{aa}
\bibliography{references.bib}

\onecolumn
\begin{appendix}

\section{Bayesian fit corner plots for each grid}

\begin{figure*}[h!]
\sidecaption
  \includegraphics[width=12cm]{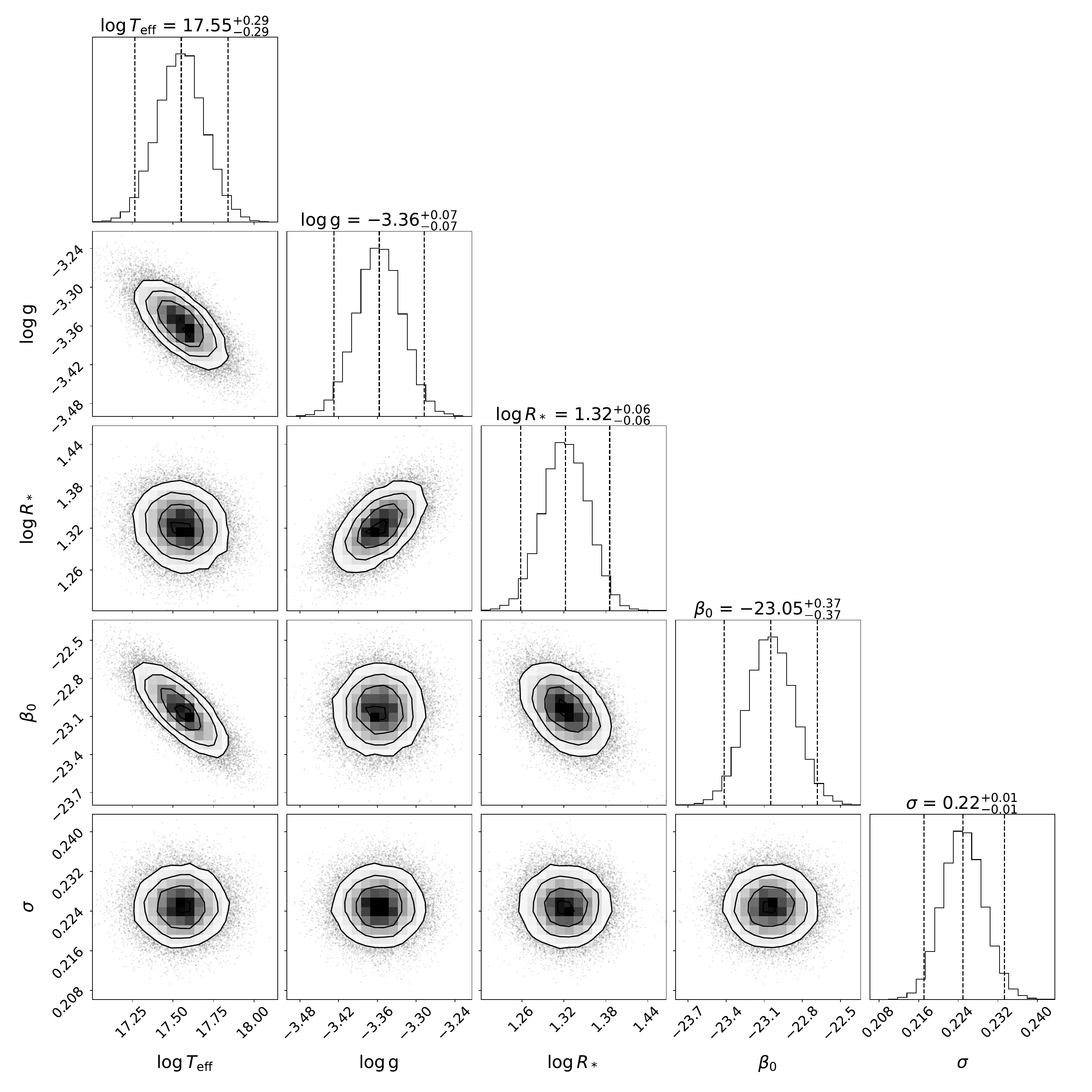}
        \caption{Results of the Bayesian linear regression performed for the mass-loss rates ($\dot{M}$) in the H grid. As observed, there are slight correlations between the adjusted temperature parameters and the intercept $\beta_0$, though these correlations are not particularly significant. Additionally, the plot highlights relatively small uncertainties in the fitted Bayesian parameters, indicating a good alignment with the chosen model. These results suggest a robust regression process, with no apparent systematic biases affecting the outcomes.\label{fig:hbayes}}        
\end{figure*}

\begin{figure*}[h!]
\sidecaption
  \includegraphics[width=12cm]{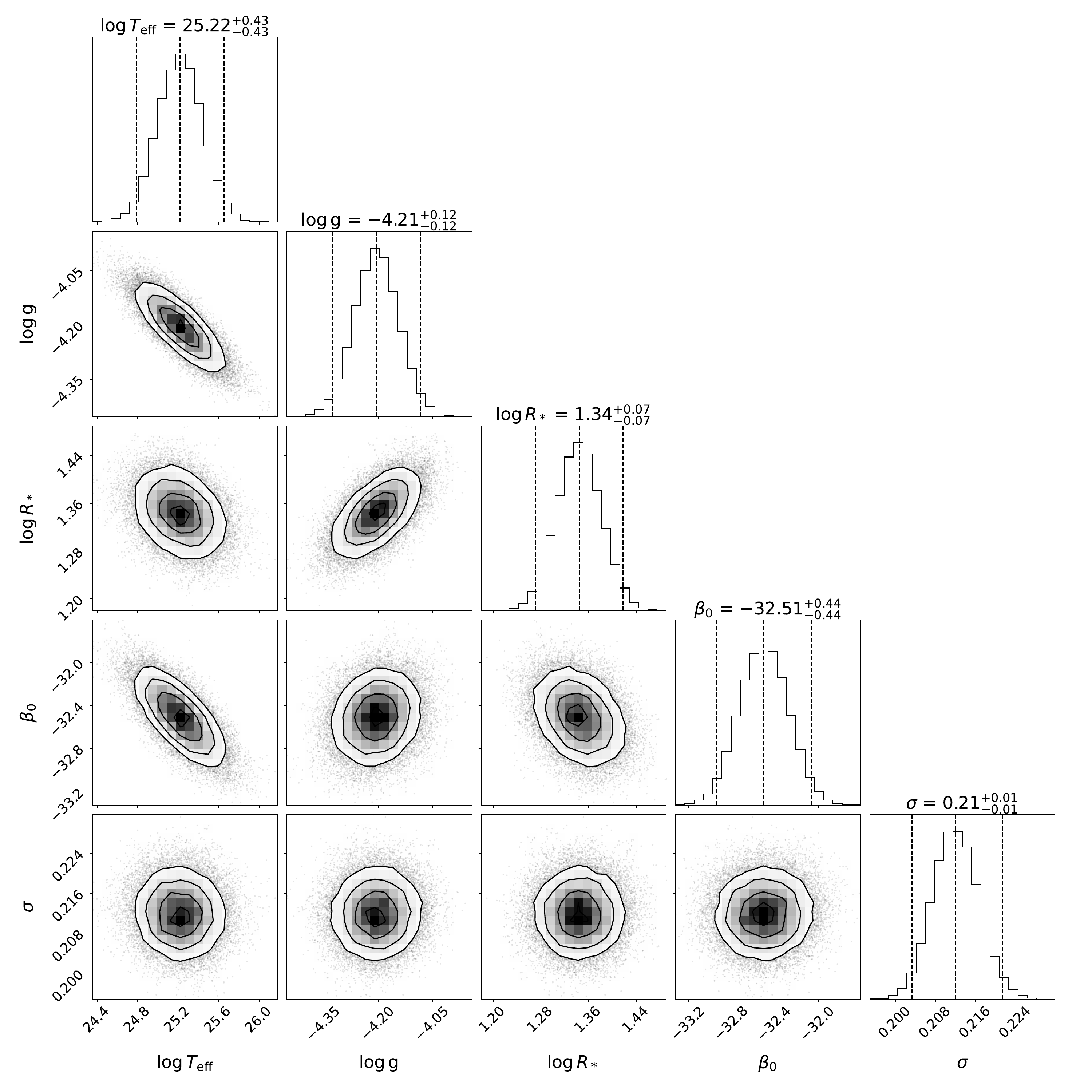}
        \caption{Corner plot showing the results of the Bayesian linear regression performed for the mass-loss rates ($\dot{M}$) in the HHe grid. The correlations follow similar trends to those seen in the H grid, maintaining low values except for $\log  T_\text{eff}$ and the adjusted parameters for $\log  g$ and $\beta_0$. Additionally, the dispersion ($\sigma$) decreases slightly compared to the H grid, which is also reflected in the Pearson's correlation coefficient ($R^2$), which increases relative to the simpler grid. 
                \label{fig:hhebayes}}   
\end{figure*}

\begin{figure*}[h!]
\sidecaption
  \includegraphics[width=12cm]{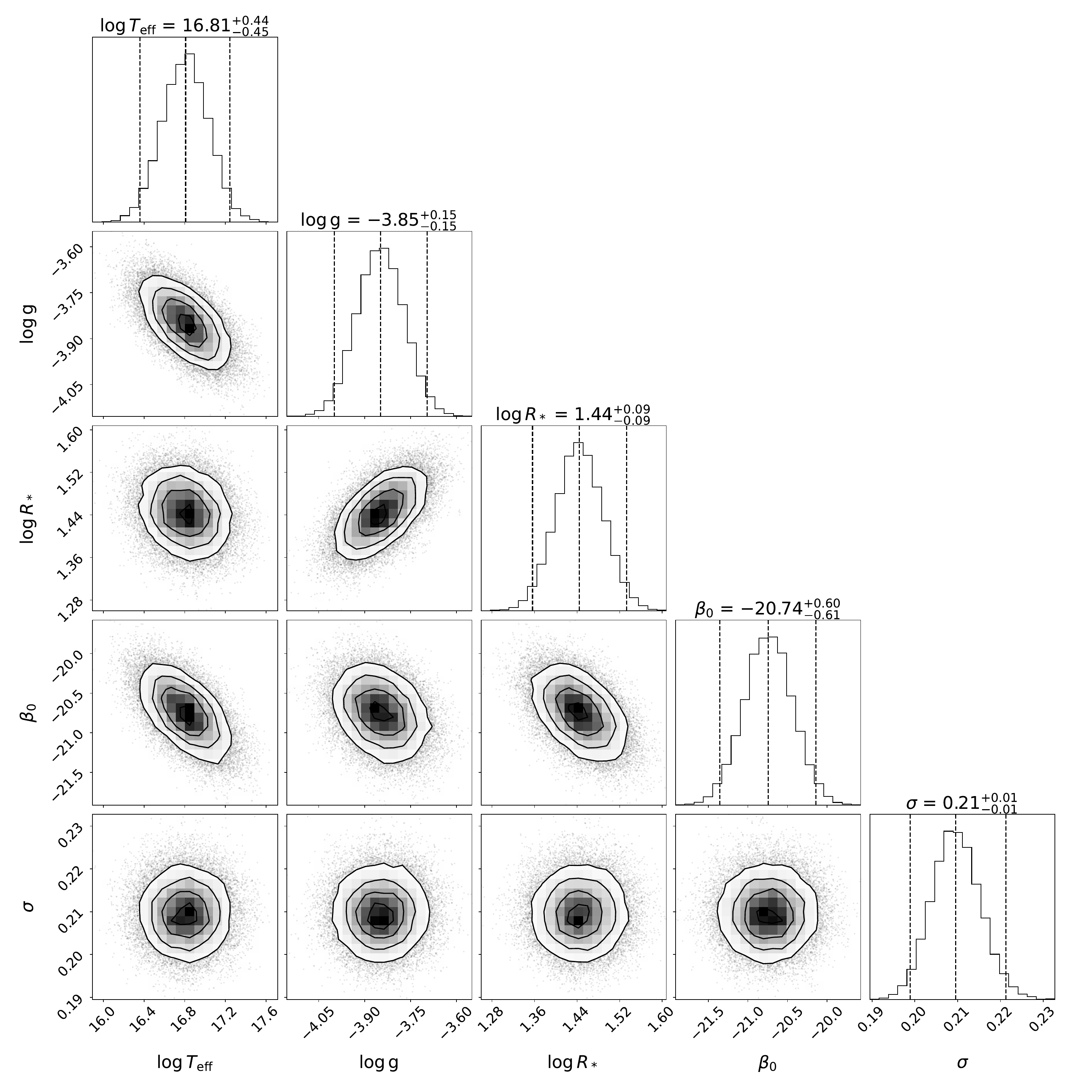}
        \caption{Corner plot showing the results of the Bayesian linear regression performed for the mass-loss rates ($\dot{M}$) in the CNO grid. As observed, the correlations remain low, except for the same parameters previously mentioned. However, the dispersion continues to follow the decreasing trend, maintaining a high $R^2$. This indicates the model performs well in describing the results obtained with our procedure, suggesting it is effective in capturing the relationships between stellar parameters and mass-loss rates.
                \label{fig:cnobayes}}   
\end{figure*}

\end{appendix}

\end{document}